# ASTROPHYSICAL S-FACTOR of p$^2$H RADIATIVE CAPTURE

S.B. Dubovichenko, A.V. Dzhazairov-Kakhramanov

*V.G. Fessenkov's Astrophysical Institute, Almaty, Kazakhstan. E-mail: albert-j@yandex.ru*

The astrophysical S-factor of p$^2$H radiative capture in the energy range down to 1 keV is considered in the potential cluster model with the classification of orbital states according to Young's scheme symmetry. It is shown that the approach used, which takes into account E1 transition only, gives a good description of the new experimental data for two potentials of the bound state of $^3$He nucleus and leads to the value S=1.35(5)·10$^{-4}$ keVb and 1.65(5)·10$^{-4}$ keVb.

The radiative p+$^2$H→$^3$He+γ capture is a part of hydrogen cycle and gives a considerable contribution to energy efficiency of thermonuclear reactions [1] which account for burning of the Sun and stars of our Universe. The interacting nuclear particles of the hydrogen cycle have a minimal value which is a potential barrier. Thus, it is the first chain of nuclear reactions which can take place at ultralow energies and star temperatures. Then, for this chain, the process of the radiative p$^2$H capture is the basic process for the transition from the primary proton fusion p+p→$^2$H+e$^-$+ν$_e$ to the final process $^3$He+$^3$He→$^4$He+2p [2] in the p-p-chain. That is why the theoretical and experimental investigation of the radiative p$^2$H capture in detail is of fundamental interest not only for nuclear astrophysics, but also for nuclear physics of ultralow energies and lightest atomic nuclei.

We will discuss the astrophysical S-factors on the basis of a potential cluster model which takes into account the supermultiplet symmetry of wave functions (WF) with the splitting of orbital states according to Young's schemes. This approach allows us to analyse the structure of inter-cluster interactions, detecting allowed and forbidden states in the interaction potential, and thus, the number of WF nodes of relative motion of clusters [3,4].

The total cross sections of the photoprocesses of the lightest nuclei were considered in this approach in our work [4]. E1 transitions resulting from the orbital part of the electric operator $Q_{Jm}(L)$ were taken into account in these calculations of the photodecays of $^3$He and $^3$H nuclei into p$^2$H and n$^2$H channels. The values of E2 cross-sections and cross-sections depending on the spin part of the electric operator turned out to be several times less. Further, it was assumed that E1 electric transitions in N$^2$H system are possible between "pure" (scheme {3}) $^2$S state of $^3$H and $^3$He nuclei and doublet $^2$P scattering state mixed according to Young's schemes {3}+{21}.

To calculate photonuclear processes in the systems under consideration the nuclear part of the potential of inter-cluster p$^2$H and n$^2$H interactions is represented as

$$V(r)=V_0\exp(-\alpha r^2)+V_1\exp(-\beta r) \quad (1)$$

with a point-like Coulomb potential, $V_0$ - the Gaussian attractive part, and $V_1$ - the exponential repulsive part.

The potential of each partial wave was constructed so that it would correctly describe the respective partial phase shift of the elastic scattering [5]. Using this concept, the potentials of the p$^2$H interaction for scattering processes were received, parameters of such potentials were fully given in works [4,6]. Then "pure" phases [3] were separated in the doublet channel and on their basis potentials of inter-cluster interaction - "pure" in accordance with Young's schemes {3} - were constructed [4,6].

The calculations of the E1 transition [4] show that the best results for the description of the total cross-sections of the $^3$He nucleus photodecay for the γ-quanta energy range 6-28 MeV, including the maximum value at Eγ=10-13 MeV, can be found if the potentials with peripheric repulsion of the $^2$P-wave of the p$^2$H scattering (table 1) and S-interaction of the bound state (BS) with parameters -34.75 MeV and 0.15 fm$^{-2}$ are used. However, this interaction gives the bound energy in the p$^2$H channel only approximately: -5.49 MeV.

The calculations of the total cross-sections of the radiative p$^2$H capture and astrophysical S-factors were done with these potentials at the energy range down to 10 keV [4]. Although, at that period of time, we only knew S-factor experimental data in the range above 150-200 keV [7]. A short time ago the new experimental data on the p$^2$H S-factor in the range down to 2.5 keV appeared in [8-10]. That is why, it is interesting to know if it is possible to describe the new data on the basis of the E1 transition in the potential cluster model with the earlier obtained P-interaction and S-potential adjusted in this work. The final parameters of $^2$S$_{\{3\}}$ and $^2$P$_{\{3\}+\{21\}}$ potentials used in the new calculations of the E1 radiative p$^2$H capture are given in [4,11] and table 1.

**Table 1.** The potentials of the p$^2$H [4] interaction in the doublet channel, used in calculations of the E1 radiative capture. E$_{BS}$ is the calculated energy of the bound state, E$_{EXP}$ its experimental value [12], {f} – Young's scheme.

| $^{2J+1}$L, {f} | $V_0$, MeV | $\alpha$, fm$^{-2}$ | $V_1$, MeV | $\beta$, fm$^{-1}$ | $E_{BS}$, MeV | $E_{EXP}$, MeV |
|---|---|---|---|---|---|---|
| $^2$S, {3} | -34.76170133 | 0.15 | --- | --- | -5.4934230 | -5.4934230 |
| $^2$P, {3}+{21} | -10.0 | 0.16 | +0.6 | 0.1 | | |
| $^2$S, {3}+{21} | -35.0 | 0.1 | --- | --- | | |

Our preliminary results [13] show that for the S-factor calculation at an energy range of about 1 keV it is necessary to improve the accuracy of finding the bound energy of the p$^2$H system in the $^3$He nucleus. It must be better than 1-2 keV. The behaviour of the tail of the wave function (WF) of the bound state (BS) should be controlled more strictly at long distances. Then, it is necessary to improve the accuracy of finding Coulomb wave functions [14] which determine the asymptotic behaviour of the scattering WF in the P-wave. For this purpose, we have rewritten our computer program, based on the finite-difference method (FDM), for calculating the total cross-sections of the E1 capture in the p$^2$H channel [14] from TurboBasic language to Fortran-90. It allowed us to essentially raise the accuracy of all calculations, including calculations of the bound energy of the $^3$He nucleus in the p$^2$H channel. Now, for example, the relative accuracy of calculating Coulomb functions, controlled by Wronskian's value, and the accuracy of finding the determinant's radical [14], which determines the accuracy of finding the bound energy, are about 10$^{-15}$.

The parameters of the "pure" doublet $^2$S-potential according to Young's scheme {3} were adjusted using these opportunities for a more accurate description of the experimental bound energy of $^3$He nuclei in p$^2$H channel. This potential has become somewhat deeper [4] and leads to a total agreement between calculated -5.4934230 MeV and experimental -5.4934230 MeV bound energies, which is obtained by using the exact mass values of particles [12]. For these computations the absolute accuracy of searching for the bound energy in our computer program was taken to be at the level of 10$^{-8}$ MeV.

The value of the $^3$He charge radius with this potential equals 2.28 fm, which is a little higher than the experimental value 1.976(15) fm [15]. The radii of proton 0.8768(69) fm and of the deuteron, 2.1402(28) fm [12] are used for these calculations and the latter is larger than the radius of the $^3$He nucleus. Thus, if the deuteron is present in the $^3$He nucleus as a cluster, it must be compressed by about 20-30% of its size in free state for a correct description of the $^3$He radius [11].

The asymptotic constant C$_W$ with Whittaker asymptotics [11,16] was calculated for controlling behavior of WF of BS at long distances; its value in the range of 5-20 fm equals C$_W$=2.333(3). The error which is given here is determined by averaging the constant in the indicated range. The experimental data known for this constant give the values 1.73-1.87 [17], which is slightly less than the value obtained here. For comparison, we can give results of three-body calculations [18], where a good agreement with the experiment [19] for the ratio of asymptotic constants for S and D waves was obtained and the value of the constant of S wave was found to be C$_S$=1.878.

In a cluster model the value of $C_W$ constant depends significantly on the width of the potential well and it is always possible to find other parameters of $^2$S-potential of BS, for example:

$V_0$=-48.04680730 MeV and $\alpha$=0.25 fm$^{-2}$, (2)
$V_0$=-41.55562462 MeV and $\alpha$=0.2 fm$^{-2}$, (3)
$V_0$=-31.20426327 MeV and $\alpha$=0.125 fm$^{-2}$, (4)

which give the same value of bound energy of $^3$He in p$^2$H channel. The potential (2) at distances of 5-20 fm leads to asymptotic constant $C_W$=1.945(3) and charge radius $R_{ch}$=2.18 fm, the variant (3) gives $C_W$=2.095(5) and $R_{ch}$=2.22 fm, the variant (4) - $C_W$=2.519(3) and $R_{ch}$=2.33 fm.

It can be seen from these results that the potential (2) with width 0.25 fm$^{-2}$ allows to obtain the most reasonable values for the charge radius and the asymptotic constant. A less deep potential may give a more accurate description of the asymptotic constant, but, as will be seen in later, will not allow us to describe the S-factor of the p$^2$H capture. In this sense, the potential (2) has the minimum acceptable width.

The variational method (VM) is used for an additional control of the accuracy of bound energy calculations for the potential from table 1, which allowed to obtain the bound energy of -5.4934228 MeV by using an independent variation of parameters and the grid having dimension 10 [14]. The asymptotic constant $C_W$ of the variational WF at distances of 5-20 fm remains at the level of 2.34(1) and the residual error doesn't exceed 10$^{-12}$ [14]. The variational parameters and expansion coefficients of the radial wave function having form

$$\Phi_L(R) = R^L \sum_i C_i \exp(-\alpha_i R^2)$$ (5)

are listed in table 2.

**Table 2.** The variational parameters and expansion coefficients of the radial WF of the bound state of the p$^2$H system for the potential from table 1. The normalisation of the function with these coefficients in the range 0-25 fm equals N=0.999999997.

| i | $C_i$ | $\alpha_i$ |
|---|---|---|
| 1 | -1.139939646617903E-001 | 2.682914012452794E-001 |
| 2 | -3.928173077162038E-003 | 1.506898472480031E-002 |
| 3 | -2.596386495718163E-004 | 8.150892061325998E-003 |
| 4 | -5.359449556198755E-002 | 4.699184204753572E-002 |
| 5 | -1.863994304088623E-002 | 2.664477374725231E-002 |
| 6 | 1.098799639286601E-003 | 4.468761998654231E+001 |
| 7 | -1.172712856304303E-001 | 8.482112461789261E-002 |
| 8 | -1.925839668633162E-001 | 1.541789664414691E-001 |
| 9 | 3.969648696293301E-003 | 1.527248552219977E-000 |
| 10 | 2.097266548250023E-003 | 6.691341326208045E-000 |

For the real bound energy in this potential it is possible to use the value -5.4934229(1) MeV with the absolute calculation error of finding the FDM energy equal to 10$^{-8}$ MeV, because the variational energy decreases as the dimension of a basis increases and gives the upper limit of the true bound energy, but the finite-difference energy increases as the size of steps decreases and the number of steps increases.

The potential (3) was examined within the frame of VM and the same bound energy of -5.4934228 MeV was received. The variational parameters and expansion coefficients of the radial wave function (5) are listed in table 3. The asymptotic constant at distances of 5-20 fm turned out to be 2.09(1) and the residual error did not exceed 2·10$^{-13}$.

The exact mass values of the particles were taken for all our calculations [12], and the $\hbar^2/m$ constant was taken to be 41.4686 MeV fm$^2$. The Coulomb parameter $\eta=\mu Z_1 Z_2 e^2/(k\hbar^2)$ was represented as $\eta = 3.44476 \cdot 10^{-2} Z_1 Z_2 \mu/k$, where k is the wave number (in fm$^{-1}$), $\mu$ the reduced mass (atomic mass unit), Z the particle charges in elementary charge units. The Coulomb potential was represented as $V_{Coul.}$(MeV) = 1.439975 $Z_1 Z_2/r$, where r is the distance (fm).

**Table 3.** The variational parameters and expansion coefficients of the radial WF of the bound state of the p$^2$H system for the potential (3). The normalisation of the function with these coefficients at the range 0-25 fm equals N=0.999999998.

| I | $C_i$ | $\alpha_i$ |
|---|---|---|
| 1 | -1.178894628072507E-001 | 3.485070088054969E-001 |
| 2 | -6.168137382276252E-003 | 1.739943603152822E-002 |
| 3 | -4.319325351926516E-004 | 8.973931554450264E-003 |
| 4 | -7.078243409099880E-002 | 5.977571392609325E-002 |
| 5 | -2.743665993408441E-002 | 3.245586616581442E-002 |
| 6 | 1.102401456221556E-003 | 5.837991732045449E+001 |
| 7 | -1.384847981550261E-001 | 1.100441373510820E-001 |
| 8 | -2.114723533577409E-001 | 2.005318455817479E-001 |
| 9 | 3.955231655325594E-003 | 1.995655373133832E-000 |
| 10 | 2.101576342365150E-003 | 8.741651544040529E-000 |

In present S-factor calculations we use the well-known formula [20]

$$S(EJ) = \sigma(EJ)E_{cm} \exp\left(\frac{31.335 Z_1 Z_2 \sqrt{\mu}}{\sqrt{E_{cm}}}\right),$$

where $\sigma$ is the total cross-section of the radiative capture process (barn), $E_{cm}$ is the center-of-mass energy of particles (keV), $\mu$ is the reduced mass (atomic mass unit) and Z are the particle charges in elementary charge units. The numerical coefficient 31.335 was received on the basis of up-to-date values of fundamental constants, which are given in [12]. The total cross-sections of radiative capture $\sigma(E)$ in a cluster model are given, for example, in the work by C. Angulo et al. [21].

In this work we considered the energy range of the radiative p$^2$H capture down to 1 keV and found the value of 1.65(5)·10$^{-4}$ keVb for the S(E1)-factor at 1 keV for the potentials from table 1. The value found is slightly lower than the known data, if we consider the total S-factor without splitting it into $S_s$ and $S_p$ parts resulting from M1 and E1 transitions. This splitting was done in [22], where $S_s(0)$=1.09(10) 10$^{-4}$ keVb and $S_p(0)$=0.73(7) 10$^{-4}$ keVb, which gives the value of 1.82(17) 10$^{-4}$ keVb for the total S-factor.

However, these are the only results with the splitting of the S-factor into M1 and E1 parts which we know and it seems that these data ought to be updated and rechecked. So, we will take as a reference point the total value of S-factor at zero energy which was measured in various works. Furthermore, the new experimental data [10] lead to the value of total S(0)=2.16(10)·10$^{-4}$ keVb and this means that contributions of M1 and E1 will change.

The known extractions of the S-factor from the experimental data, without splitting to M1 and E1 parts, at zero energy give the value of 1.66(14) 10$^{-4}$ keVb [23]. The previous measurements by the same authors gave 1.21(12)10$^{-4}$ keVb [24] and the value 1.85(5)10$^{-4}$ keVb was received in [25]. The average of these experimental measurements equals 1.69(58) 10$^{-4}$ keVb what is in a good agreement with the value 1.65(5) 10$^{-4}$ keVb calculated here only on the basis of the E1 transition.

Our calculation results for the S-factor of the p$^2$H capture with the potentials from table 1 at the energy range from 1 keV to 10 MeV are shown in figs. 1 and 2 by dotted lines and at energies above 10 keV there are practically no differences from our previous results [4]. Now the calculated S-factor reproduces experimental data at the energies down to 10-20 keV comparatively well and at lower energies the calculated curve practically falls within the experimental error band of the work [10].

Solid lines in figs. 1 and 2 show the results for potential (3) which describes the behavior of the S-factor somewhat better at energies from 50 keV to 10 MeV and which gives the value of S=1.35(5)·$10^{-4}$ keVb for the energy of 1 keV. At energies of 20-50 keV the calculation curve follows the line of the lower limit of the error band of work [9], and at energies below 10 keV it falls within the experimental error band [10].

The dashed lines in figs. 1 and 2 show the results for potential (4) and the dash-dotted line those for potential (2). Potential (2) with the asymptotic constant 1.945, which is the closest to the experimental value, allows us only to describe correctly the S-factor within the range from 50 keV to 3 MeV. At the energy of 2.5 keV it leads to the results which fall within the error band of work [10] and at 1 keV gives a value of the S-factor equal to 1.15(5)·$10^{-4}$ keVb, which is also within the experimental error band - 1.7(6) $10^{-4}$ keVb. At the same time, potential (4) with the overestimated asymptotic constant of 1.15(5)·$10^{-4}$ keVb completely describes the new data [10] below 20-30 keV and at the energy of 1 keV it gives the S-factor value 1.88(5)·$10^{-4}$ keVb which is in better agreement with the results [22,25].

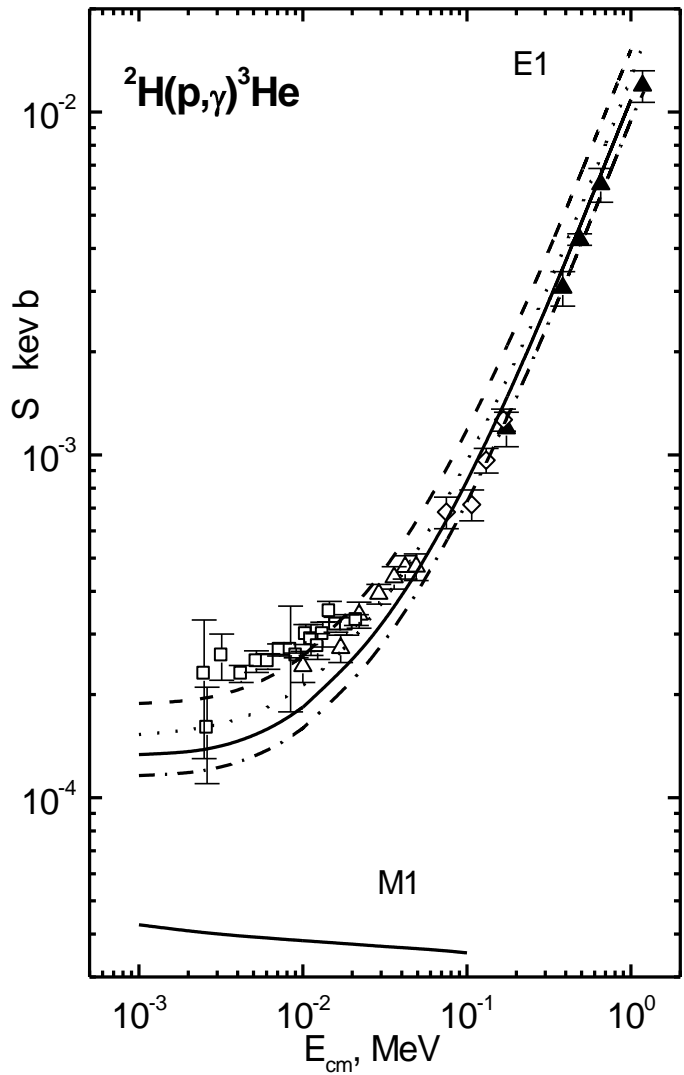

**Fig. 1.** Astrophysical S-factor of p$^2$H radiative capture in the range 1 keV-1 MeV. Lines: calculations with the potentials mentioned in the text. Triangles denote the experimental data from [7], open rhombs from [8], open triangles from [9], open blocks from [10].

From these calculations one may conclude that the best results are obtained with the BS potential (3) which describes the experimental data in the widest energy range and which could be considered as a revised version of our previous potential shown in table 1. It represents a sort of a compromise in describing asymptotic constant (2.095), charge radius (2.22 fm) and S-factor of the radiative $p^2H$ capture within the whole range of considered energies.

The M1 transition from the S scattering state, which is mixed in accordance with Young's schemes, to the bound state, which is "pure" according with orbital symmetries of the S state of the $^3He$ nucleus, can give a contribution at low energies. For our calculations we used the doublet S-potential of the scattering states with the parameters listed in table 1 and the BS potential (3). The calculation results at the energies 1-100 keV are shown in fig. 1 by the solid line at the bottom of the figure. It can be seen that the cross-section of the M1 process is several times lower than the cross-section of the E1 transition.

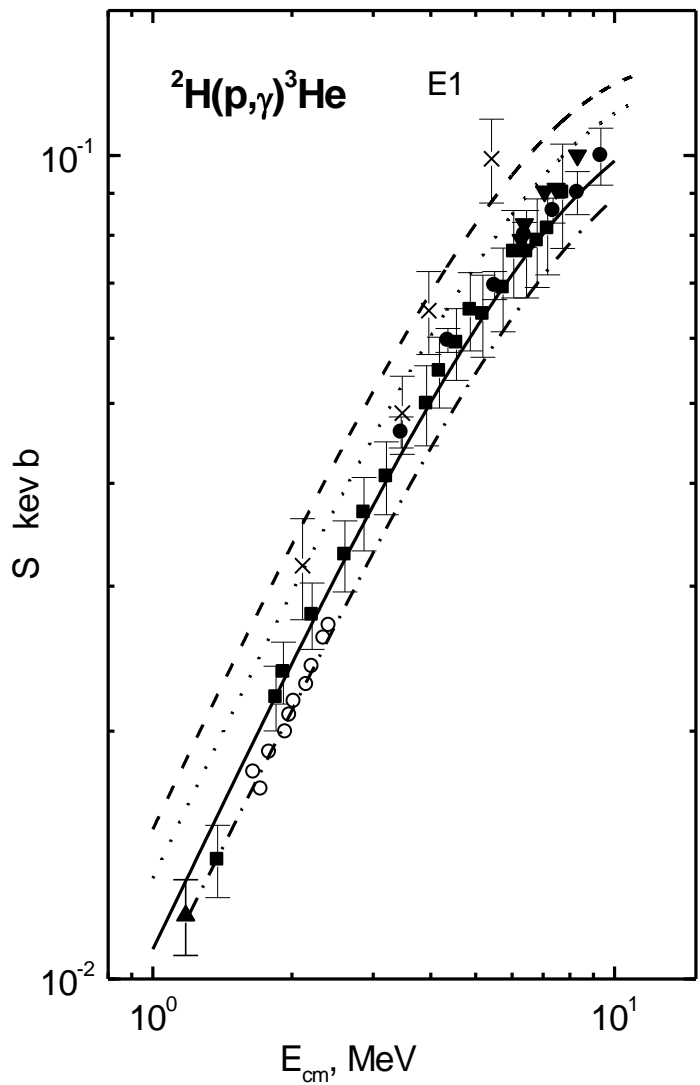

**Fig. 2.** Astrophysical S-factor of $p^2H$ radiative capture in the range 1 MeV-10 MeV. Lines: calculations with the potentials mentioned in the text. Triangles denote the experimental data from [7], squares are from work [26], black points from [27], crosses from [28], inverted triangles from [29], open circles from [30].

However, it is necessary to note that we are unable to build the scattering S-potential uniquely because of the ambiguities in the results of different phase shift analyses. The other variant of potential with parameters $V_0$=-55.0 MeV and $\alpha$=0.2 $fm^{-2}$ [31], which also describes well the S phase shift, leads at these energies to cross-sections of the M1 process several times higher than those of E1. Thus, such a big ambiguity in parameters of the S-potential, associated with errors of scattering phase shifts extracted from the experimental data, does not allow us to make certain conclusions about the contribution of the M1 process in the p$^2$H radiative capture.

The BS potentials are defined by the bound energy, asymptotic constant and charge radius quite uniquely. The potential description of the scattering phase shifts, which are "pure" in accordance with Young's schemes, is the additional criteria for determination of such parameters. Then, for the construction of the scattering potential it is necessary to carry out a more accurate phase shift analysis for the $^2$S-wave and to take into account the spin-orbital splitting of $^2$P phase shifts at low energies, as was done for the elastic p$^{12}$C scattering at energies 0.2-1.2 MeV [32]. This will allow us to adjust the potential parameters used in the calculations of the p$^2$H capture in the potential cluster model, the results of the calculations of which depend strongly on the accuracy of the construction of the interaction potentials according with the scattering phase shifts.

Thus, the S-factor calculations of the p$^2$H radiative capture for the E1 transition at the energy range down to 10 keV, which we carried out about 15 years ago [4] when only the experimental data above 150-200 keV were known, are in a good agreement with the new data of works [8,9] in the energy range 10-150 keV. Therefore, the potential cluster model with forbidden states taking into account Young's scheme symmetry turned out to be able to give, in general, a correct behaviour prediction of the S-factor of the p$^2$H capture at energies down to 10-20 keV [4,31].

The calculations of the S(E1)-factor at the lower energy range show that it tends to remain constant at energies 1-3 keV. The new results, including the ones for potential (3) at the energies lower than 10 keV, practically fall within the error band of work [10], where the S-factor was measured at an energy range down to 2.5 keV.